\documentstyle[aps,multicol,psfig]{revtex}
\newcommand{\beq}{\begin{equation}}
\newcommand{\eeq}{\end{equation}}
\newcommand{\beqa}{\begin{eqnarray}}
\newcommand{\eeqa}{\end{eqnarray}}
\def\ket#1{|\,#1\,\rangle}

\def\braket#1#2{\langle\, #1\,|\,#2\,\rangle}

\def\opone{\leavevmode\hbox{\small1\kern-3.8pt\normalsize1}}

\begin{document}
\title{Quantum Key Distribution using Multilevel Encoding: Security Analysis}
\author{Mohamed Bourennane\thanks{Electronic address: boure@ele.kth.se}, Anders Karlsson, and Gunnar Bj\"{o}rk}
\address{Department of Microelectronics and Information Technology,\\ 
Royal Institute of Technology (KTH), Electrum 229, SE-164 40 Kista, Sweden}

\author{Nicolas Gisin}
\address{GAP-Optique, Universite de Geneve, 20 rue de \\ l'Ecole de Medecine, 1211 Geneve 4, Switzerland}

\author{Nicolas J. Cerf}
\address{Ecole Polytechnique, CP 165, Universit\'e Libre de Bruxelles,
B-1050 Bruxelles, Belgium \\ 
and Jet Propulsion Laboratory,
California Institute of Technology, Pasadena, CA 91109}

\date{\today}
\draft
\maketitle

\begin{abstract}
We present security proofs for a protocol for Quantum Key Distribution (QKD) based on encoding in finite high-dimensional Hilbert spaces. This protocol is an extension of Bennett's and Brassard's basic protocol from two bases, two state encoding to a  multi bases, multi state encoding. We analyze the mutual information  between the legitimate parties and the eavesdropper, and the error rate, as function of the dimension of the Hilbert space, while considering optimal incoherent and coherent eavesdropping attacks. We obtain the upper limit for the legitimate party error rate to ensure unconditional security when the eavesdropper uses incoherent and coherent eavesdropping strategies. We have also consider realistic noise caused by detector's noise. 

\end{abstract}

%\pacs{03.67.Dd, 03.67.Hk, 03.67.-a, 03.65.Bz}

\vspace{-0.4cm}

%\begin{multicols}{2}
\section{Introduction} 
Quantum cryptography aims to provide an {\em unconditionally} secure key distribution
between two parties, Alice and Bob. In the first protocol proposed by Bennett and Brassard (BB84) \cite{BB84}, to detect eavesdropping, Alice and Bob choose randomly between two complementary (conjugate) bases and in each basis the \textquotedblleft information\textquotedblright\ is encoded using  two orthogonal quantum states (qubits).  Since the basis is unknown to the eavesdropper (by convention called Eve), she cannot simply copy the sent states because the non-cloning theorem. The use of random complementary bases furthermore implies that if the sender Alice prepares a state in one basis, the outcome of a measurement by Bob or Eve in a complementary basis will yield a totally random measurement outcome. These features guarantee that any eavesdropping  attempt will invariably introduce errors in the transmission, which can be detected by the legitimate communicating parties. An extension to the BB84 protocol was made by Bru\ss\ \cite{DBruss} and by Bechmann-Pasquinucci and Gisin \cite{HBech-NGisin} to a six-state, three complementary bases protocol. The analysis shows that Eve's information gain for a given impaired error rate is lower than in the BB84 protocol \cite{DBruss,HBech-NGisin}. Very recently two other extensions were proposed where the authors have considered schemes using four states and two bases \cite{HBech-WTittel}, and three states and four bases \cite{HBech-APeres}. In an earlier work we have generalized these results to encoding in $N$-dimensional Hilbert space using $M \leq N+1$ bases \cite{Bourennane} where we have considered some specific and rather simple, but realistic, eavesdropping attacks. The goal of this  work is to find an ultimate and practical condition for the security of quantum key distribution protocols, sufficiently general to encompass all possible types of eavesdropping. The condition we derive is given in the form of a theorem. We will also derive the upper permissible limit for Bob's error rate to ensure unconditional security when Eve uses incoherent and coherent eavesdropping attacks. 

The paper is organized as follows: In Sec. 2, we  give a brief introduction to our protocol. In Sec. 3 we reiterate the secrecy capacity of a channel and derive the results for an intercept-resend eavesdropping attack. In Secs. 4 and 5, we study optimal individual eavesdropping attacks, and coherent eavesdropping attacks, respectively. In Sec. 6 we consider realistic systems where we assume that the detector dark count probability is not negligible. Finally, in Sec. 7 we present our conclusions.

%----------------------------------------------------------------------

\section{A Multi Bases Multi State Quantum Key Distribution Protocol}
In the BB84 protocol \cite{BB84}, Alice first randomly chooses between one of two bases to prepare her state, and secondly she randomly decides which of two orthogonal states in the chosen basis to send. Extending this protocol to a $N$-dimensional Hilbert space ${\cal H}_N$, Alice first chooses from which of $M$ complementary bases to choose her state from, and secondly she decides which of the $N$ orthogonal states defining the basis to send. The \textquotedblleft information\textquotedblright\ encoded by the chosen state will from hereon be denoted quNits.
Each symbol sent in the $M$ bases and $N$ quNits are chosen randomly with equal probability, i.e. each of the possible $NM$ states appear with probability $1/(MN)$.
We first define the bases $\{{\psi}_A\}$ and   $\{{\psi}_B\}$ over an $N$-dimensional space to be {\em mutually complementary} if the inner products between all possible pairs of vectors, with one state from each basis, have the same magnitude:
\begin{eqnarray}
|_A \braket{{\psi}_{i}}{{\psi}_{j}}_B |=1/\sqrt{N} \qquad \forall \quad i,j. 
\label{overlap}
\end{eqnarray}
If a quantum state is prepared in the $\{\psi_A\}$ basis, but measured
in the complementary $\{\psi_B\}$ basis, the outcome is completely random.
Wootters and Fields have shown \cite{Wootters-Fields} that when $N= p^k$, where $p$ is a
prime and $k$ a positive integer, which we restrict ourselves to here, then there exist a set of $M=N+1$ mutually complementary bases \cite{Wootters-Fields}.

To estimate the mutual information between Alice and Bob, Alice and Eve, and the information gain of the eavesdropper Eve, the  relevant information measure  is the Shannon information of the \textit{sifted} symbols, i.e., the symbols for which Alice and Bob have used the same bases. For simplicity, we choose to measure this information in bits.  From the receiver's (Bob's or Eve's) point of view, there will be   an \textit{a priori}  $p(x)$ and an \textit{a posteriori} $p(x\vert y)$ probability, the latter being the conditional probability of the sending party (Alice) having sent the symbol $x$, given that the receiver (Eve or Bob) measured the result $y$.  The receiver's mean information gain from Alice's symbol, $I_{AY}^N$, where $Y= B, E$ denote either Bob or Eve, equals his or her entropy decrease:
\begin{eqnarray}
I_{AY}^N  = H_{\textit{apri}}^N - H_{\textit{apost}}^N .
\label{Shannon1}   
\end{eqnarray}
The \textit{a priori} probability for Alice's symbol is uniform (since the protocol dictates that Alice must chose the symbols she sends randomly), leading  to $H_{\textit{apri}}^N = \log(N)$. The \textit{a posteriori} entropy is defined: 
\begin{eqnarray}
 H_{\textit{apost}}^N = \sum_y p(y)\sum_x p(x\vert y)\log(p(x \vert y)),
\label{Shannon2}   
\end{eqnarray}
where the \textit{a posteriori} probability of symbol $y$ given observer's result $x$ is given by Bayes' theorem:
\begin{eqnarray}
p(x \vert y)= \frac{p(y \vert x)p(x)}{p(y)},
\label{Shannon3}   
\end{eqnarray}
with $p(y) = \sum_x p(y \vert x)p(x)$. 

The mutual information between Bob and Alice as function of Bob's error rate is obtained by using Eq. (\ref{Shannon1}) and by using the symmetry properties of the protocol, i.e., that Bob's measurement errors are independent of, and uniform for, all symbols sent by Alice:
\begin{eqnarray}
I_{AB}^N(e^N_B)= \log(N) + (1-{e^N_{B}})\log(1-e^N_{B}) + e^N_{B}\log(\frac{e^N_{B}}{N-1}),
\label{eq: info}
\end{eqnarray}
where $e_{B}^N$ is Bob's error rate, i.e., the probability that he measures a symbol erroneously. Note that since expressions (\ref{Shannon1} ) and (\ref{eq: info}) refer to the information and errors contained in the sifted symbols, these errors are due to  a possible eavesdropping disturbance and system noise such as the dark counts of the detectors, the tranmission loss, etc. They are not due to Bob's random choice of measurement basis.
%------------------------------------------------------------------------------
\par
\section{Eavesdropping}
In an ideal system, after the quNit string has been transmitted, measured, and sifted, Alice and Bob will share a common key. However, in real systems there are always some errors, and some of these errors may be due to an eavesdropper. Hence, Alice and Bob need to use error correction through a classical channel to establish an identical key, and privacy amplification to obtain a secret common key \cite{error,privacy}. 
The eavesdropping attacks by Eve must introduce errors. As stated above this is due to Alice's random choice of measurement basis and the fact that Eve cannot copy an unknown state perfectly.  In the case of simple intercept-resend eavesdropping attacks, Eve gets one of the $NM$ possible results. After Alice and Bob announced their choice of bases we have $p(x =y | \{ \psi_A\} = \{ \psi_E\} ) = 1$, $p(x \neq y | \{ \psi_A\} = \{ \psi_E\} ) = 0$, and $p(y | \{ \psi_A\} \neq \{ \psi_E\} )= 1/N$ $\forall$ $x,y$. Therefore, according to (\ref{Shannon1}) and (\ref{eq: info}), Eve's information gain is $I_{AE} = \log(N)/M$ and Bob's error rate becomes $e_{B}^N = (1-1/M)(1-1/N)$ when Eve employs the intercept-resend eavesdropping strategy. 

Csisz\'{a}r and K\"{o}rner \cite{Csiszar} have given a lower bound for
the {\em secrecy capacity}, that is, the maximum rate at which
Alice can reliably send random symbols to Bob such that the rate at which Eve obtain
information about the symbols is arbitrarily small. We can give their result as a theorem, the proof of the theorem is given in \cite{Csiszar}.\\

{\em Theorem 1: } Alice and Bob can establish a secret key (using error correction and privacy amplification) if, and only if, $I_{AB}^N \geq I_{AE}^N$ or $I_{AB}^N \geq I_{BE}^N$, where $I_{AB}^N$, $I_{AE}^N$ and $I_{BE}^N$ are the mutual information between Alice and Bob, Alice and Eve, and Bob and Eve, respectively.\\

Taking the sifting, error correction, and privacy amplification  into account, we can hence define an effective transmission rate as
\begin{eqnarray}
R_{AB}^N(e^N_B) = \frac{1}{M}(I_{AB}^N(e^N_B) - I_{AE}^N(e^N_B)). \label{RAB}
\end{eqnarray}

We will discuss in the following sections the different eavesdropping strategies and present a security analysis. First we begin by considering individual attacks where Eve attaches independent probes to each quNit and measures her probes one after the other. Second, we consider coherent attacks in which Eve process several quNits jointly. 
%------------------------------------------------------------------------
\section{Individual Eavesdropping Attacks: Universal Quantum Cloning Machine}
%---------------------------------------------------------
\par
Here we discuss an individual eavesdropping strategy
based on the use of an asymmetric version of the N-dimensional symmetric
Universal Quantum Cloning Machine  (UQCM) introduced
by Bu\v{z}ek and Hillery \cite{Buzek1}. This asymmetric cloner \cite{CERF0,CERF}
can be used to obtain two copies of Alice's quantum state that are not of the
same fidelity. Eve then keeps one of the copies (typically, the bad one)
for herself, and passes the other copy  (typically, the good one) to Bob. 
Then, after Bob and Alice have announced their chosen bases, Eve does the same measurement as
Bob did, i.e., she measures her copy in the same basis as Alice and Bob. Therefore, on the average, she will obtain similar information as Bob. The asymmetry parameter of the cloner 
allows her to adjust the amout of information she gained, and thereby the amount of
information Bob lost. The asymmetric cloner is universal, just as the UQCM \cite{Buzek1}, 
so that all input states are copied equally well (Bob's fidelity and Eve's fidelity
do not depend on Alice's chosen state nor chosen basis). Note that the quantum circuit that
implements this asymmetric cloner is exhibited in Ref.~\cite{Buzek2}.

It should be stressed that, if Eve makes 
use of such an asymmetric UQCM for eavesdropping, she should exploit the state of her copy
but also that of the cloning machine (or ancilla) in order to extract a maximum of
information on Alice's quantum state. In particular,
she can make a coherent measurement on the state of the cloning machine and her copy 
in order to infer whether she introduced an error at Bob's station (and precisely what error) \cite{HBech-NGisin}.
For increasing disturbance, the fidelity $F^N_{AB}$ between the sent state and the state inferred by Bob (defined on the sifted symbols) that govern the probability that he and Alice will accept the transmitted state decreases, while Eve's probability of correctly guessing the symbol  increases.

Let us analyze the situation when Eve uses a $N$-dimensional copying machine
such as described in Re. \cite{CERF0,CERF}. 
If Alice sends the state $\ket{\psi_k}$, the output state is given by:
\begin{equation}  \label{nic1}
|\psi_{k}\rangle_{A} \rightarrow \sum_{m,n=0}^{N-1} a_{m,n} U_{m,n}|\psi_{k}\rangle_{B}|\Psi_{m,N-n}\rangle_{EM}
\end{equation}
where the amplitudes $a_{m,n}$ (with $m,n=0,\cdots, N-1$) characterize the cloner and A, B, E and M stand for Alice, Bob, Eve and the cloning machine respectively. 
Here, the states $|\Psi_{m,n}\rangle_{EM}$ are the generalization the Bell states, 
that is, a set of $N^2$ orthonormal maximally-entangled states 
of two $N$-dimensional systems:
\begin{equation}
|\Psi_{m,n}\rangle_{EM} = {1\over\sqrt{N}} \sum_{l=0}^{N-1}
{\rm e}^{2\pi i (l n / N)} |\psi_l\rangle_E |\psi_{l+m} \rangle_M 
\end{equation}
where the indices $m$ and $n$ ($m,n=0,\cdots, N-1$) 
label the $N^2$ states. Note that, here and below, the ket labels
are taken modulo $N$. 
%It is easy to check that the $|\psi_{m,n}\rangle$ are orthonormal
%and form a complete basis in the product Hilbert spaces
%${\cal H}_A\otimes {\cal H}_B$. 
The operators $U_{m,n}$, defined as
\begin{equation}
U_{m,n} = \sum_{k=0}^{N-1} {\rm e}^{2\pi i (k n / N)}
|\psi_{k+m}\rangle\langle \psi_{k}|
\end{equation}
form a group of error operators on $N$-dimensional states, generalizing the Pauli matrices
for qubits: $m$ labels the ``shift'' errors (generalizing the bit flip $\sigma_x$)
while $n$ labels the phase errors (generalizing the phase flip $\sigma_z$).
Using the definition of the states $|\Psi_{m,n}\rangle$ and operators $U_{m,n}$,
Eq.~(\ref{nic1}) can be reexpressed as
\begin{equation}
|\psi_{k}\rangle_{A} \rightarrow {1\over \sqrt{N}} \sum_{m=0}^{N-1} |\psi_{k+m}\rangle_B
\sum_{l=0}^{N-1} c_{m,k-l} \, |\psi_l\rangle_E \, |\psi_{l+m}\rangle_M 
\end{equation}
where
\begin{equation}
c_{m,j} = \sum_{n=0}^{N-1} a_{m,n} {\rm e}^{2\pi i (j n / N)}
\end{equation}
which will be used in the following.
Tracing the output joint state as given by Eq.~(\ref{nic1})
over $E$ and $M$, it is easy to check that Alice's state
$|\psi_{k}\rangle_{A}$ gets transformed, at Bob's station, into the mixture
\begin{equation}
\rho_B = \sum_{m,n=0}^{N-1} |a_{m,n}|^2 |\psi_{k+m}\rangle\langle\psi_{k+m}|
\end{equation}
Thus, the state undergoes a $U_{m,n}$ error with probability 
$p_{m,n}=|a_{m,n}|^2$ (with $\sum_{m,n} p_{m,n}=1$).
Note that $U_{0,0}=\openone$, implying that the state is left
unchanged with probability $p_{0,0}$. The phase errors $(n\ne 0)$ clearly do not play
any role in the above mixture, so the fidelity for Bob can be expressed as
\begin{equation}
F_B = \langle \psi_k|\rho_B|\psi_k\rangle
= \sum_{n=0}^{N-1} |a_{0,n}|^2
\end{equation}

Now, we will impose that the cloner descibed above is universal, that is \cite{CERF0,CERF}
\begin{equation}
a_{m,n} = \alpha \, \delta_{m,0} \, \delta_{n,0} + \frac{\beta}{N}
\end{equation}
with the normalisation relation
\begin{equation}
\alpha^2  +\frac{2}{N}\alpha\beta + \beta^2 = 1.
\end{equation}
where the balance $\alpha$ vs $\beta$ parametrizes the asymmetry of the cloner
($\alpha=1$ and $\beta=0$ correspond to the case where Bob gets all the information,
whereas $\alpha=0$ and $\beta=1$ correspond to Eve getting all the information).
This implies that
\begin{equation}
c_{m,j} = \alpha \, \delta_{m,0} + \beta \, \delta_{j,0}
\end{equation}
so that we obtain for the cloning transformation
%Using the above relations and the following relation
%\begin{equation}
%\sum_{n=0}^{N-1} {\rm e}^{2\pi i [(j-j') n / N]} = N \, \delta_{j,j'}
%\end{equation}
%After some simple algebra, it is easy to obtain 
\begin{eqnarray}  \label{nic2}
|\psi_{k}\rangle_{A} &\rightarrow& 
{1\over\sqrt{N}}
\sum_{m=0}^{N-1} |\psi_{k+m}\rangle_{B}
\left( \alpha\,\delta_{m,0} \sum_{l=0}^{N-1} |\psi_{l}\rangle_{E} |\psi_{l}\rangle_{M}  
+\beta |\psi_{k}\rangle_{E} |\psi_{k+m}\rangle_{M} \right)    \nonumber \\
&=& |\psi_{k}\rangle_{B} \left( 
\frac{\alpha}{\sqrt{N}} \sum_{l=0}^{N-1} |\psi_{l}\rangle_{E} |\psi_{l}\rangle_{M} 
+ \frac{\beta}{\sqrt{N}} |\psi_{k}\rangle_{E} |\psi_{k}\rangle_{M} \right)\nonumber \\ 
&+& \sum_{m=1}^{N-1} |\psi_{k+m}\rangle_B \left( \frac{\beta}{\sqrt{N}} |\psi_{k}\rangle_{E} |\psi_{k+m}\rangle_{M} \right)   
\end{eqnarray}
where the first term in the r. h. s. corresponds to Bob having no error, while the $(N-1)$
other terms correspond to all possible errors for Bob.

Eve' strategy is as follows. She first measure both her copy $E$ and the ``cloning
machine'' M in the good basis (after the chosen basis is disclosed by Alice and Bob).
If the two outcomes coincide, then she knows for sure that Bob has no error 
($m=0$), so that the state she has is the first term in the r. h. s. 
of Eq.~(\ref{nic2}). Otherwise, she knows Bob had an error ($m>0$), and she gets one of the
other terms in the r. h. s. of Eq.~(\ref{nic2}). Let us consider these two cases:
\\
(i) $m = 0$. The joint probability that Eve obtains $m=0$ and the right value of $k$ is
\begin{equation}
p_{m = 0}(k) = \frac{(\alpha + \beta)^2}{N} 
\end{equation} 
while the probability that she obtains $m=0$ with any other of the $(N-1)$ possibilities 
$l \neq k$ is
\begin{equation}
p_{m = 0}(l) = \frac{\alpha^2}{N} 
\end{equation} 
(ii) $m \neq 0$. Then, a measurement of her copy gives Eve the right value of $k$ 
with certainty. Thus, the joint probability that Eve obtains any of the $N-1$ values 
of $m\neq 0$ together with the good $k$ is
\begin{equation}
p_{m \neq 0}(k) =  \frac{\beta^2}{N} 
\end{equation} 
The fidelity of Bob is given by
\begin{eqnarray}
F_B &=& p_{m = 0}(k) +\sum_{l\neq k} p_{m = 0}(l) \nonumber \\
&=& \frac{(\alpha + \beta)^2}{N} + (N-1)\frac{\alpha^2}{N} \nonumber \\
&=& 1 -\frac{N-1}{N} \beta^2
\end{eqnarray} 
The corresponding mutual information between Alice and Bob is given by
\begin{eqnarray}
I(A{\rm:}B)  &=& \log(N) - H\left[F_B, \frac{1-F_B}{N-1},..., \frac{1-F_B}{N-1}\right] \nonumber \\
&=& \log(N) + F_B\log\left[F_B\right] + (1-F_B)\log\left[\frac{1-F_B}{N-1}\right].
\end{eqnarray} 
Consider now the mutual information between Alice and Eve. Conditionally on
Eve's measured value of $m$ (i.e., conditionally on Bob's error), this information
can be expressed as 
\begin{eqnarray}
I(A{\rm:}E|m = 0)  &=& \log(N) - H\left[\frac{(\alpha+ \beta)^2}{NF_B}, \frac{\alpha^2}{NF_B},..., \frac{\alpha^2}{NF_B}\right]  \nonumber \\
I(A{\rm:}E|m \neq 0)  &=& \log(N) 
\end{eqnarray} 
Thus, the average mutual information between Alice and Eve is 
\begin{eqnarray}
I(A{\rm:}E)  &=& F_B \, I(A{\rm:}E| m = 0) + (1-F_B) \, I(A{\rm:}E|m \neq 0)  \nonumber \\
&=& \log(N) - F_B \, H\left[\frac{(\alpha+ \beta)^2}{NF_B}, \frac{\alpha^2}{NF_B},..., \frac{\alpha^2}{NF_B}\right] \nonumber \\
&=& \log(N) + \frac{(\alpha + \beta)^2}{N}\log\left[\frac{(\alpha + \beta)^2}{NF_B}\right] + \frac{N - 1}{N}\alpha^2\log\left[\frac{\alpha^2}{NF_B}\right].
\end{eqnarray} 
This information can also be reexpressed, using Eve's fidelity,
\begin{equation}
F_E = 1 -\frac{N-1}{N} \alpha^2
\end{equation}
as
\begin{eqnarray}
I(A{\rm:}E) 
&=& \log(N) - F_B \, H\left[\frac{F_B+F_E-1}{F_B}, \frac{1-F_E}{(N-1) F_B},..., \frac{1-F_E}{(N-1) F_B}\right] \nonumber \\
&=& \log(N) + (F_B+F_E-1) \log\left[\frac{F_B+F_E-1}{F_B}\right] + 
(1-F_E) \log\left[\frac{1-F_E}{(N-1) F_B}\right].
\end{eqnarray}

As shown in \cite{Buzek1} the maximal fidelity of copying a quNit is obtained using the UQCM. This  maximal value of the fidelity correspond precisely to the fidelity of optimal incoherent eavesdropping strategy, as Bechmann-Pasquinucci and Gisin have shown explicitly for the ($N=2,M=3$) case \cite{HBech-NGisin}.
From the symmetry of the problem it follows that for $M=N+1$, the fidelity of the optimal incoherent eavesdropping is accomplished using a UQCM.

In Fig. \ref{fig:IABN}, we plot the information rate $R_{AB}^N$, defined by Eq. (\ref{RAB}), as a function of Bob's error rate for different values of $N$. For each $N$, the intersection between the graphs $R_{AB}^N$ and the horizontal axis correspond to the upper permissible bound for Bob's error rate to enable secure key distribution. In all cases, the UQCM gives the best performance (from the viewpoint of the eavesdropper) so Alice and Bob should use the UQCM model to estimate the \textquotedblleft leakage\textquotedblright\ of information to Eve when applying privacy amplification. 

%---------------------------------------------------------------------------
\section{Coherent Eavesdropping Attacks}
In the previous section we have assumed only individual attacks, i.e., that Eve  manipulates and performs measurements on each quNit separately. In this section we address the question: If Eve manipulates several quNits coherently, means an arbitrary large but finite number of quNits. We like to stress that the length of key must be much longer than this number.  What is the maximum rate of errors detected by Bob that allows Alice and Bob to still apply error correction and privacy amplification to extract a secure key? Already in 1996, Mayers presented ideas on how to prove this bound \cite{Mayers1}. Now several proofs exist \cite{Mayers1,Lo,Biham,Shor}.
Here we shall present the proof in form quite different from the previous ones. We present our result in the form of a theorem:\\

{\em Theorem 2:}
In  $N$-dimensional Hilbert space, two users Alice and Bob can establish a secret key if, and only if, Bob's error rate satisfy the inequality: \begin{eqnarray}
(1-e^N_{B})\log(1-e^N_{B}) + e^N_{B}\log(\frac{e^N_{B}}{N-1}) \leq -\frac{1}{2}\log(N) ,
\label{ineq4}
\end{eqnarray}
where $e_{B}^{N}$ is Bob's error rate.

To prove this theorem we need another theorem due to Hall \cite{Hall} that sets a limit on the sum of the mutual information between Alice and Bob and the mutual information between Alice and Eve:\\

{\em Theorem 3:}\\
Let $\hat{B}$ and $\hat{E}$ be symbol observables for Bob and Eve, respectively, in a $N$-dimensional Hilbert space such that the  maximum possible overlap between any two eigenvectors $\ket{\psi_i}_B$ and $\ket{\psi_j}_E$ corresponding to these observables is $C = \textrm{Max}_{\, i, j} \{| _B\braket{\psi_i}{\psi_j}_E | \}$. Then the mutual information Alice-Bob and Alice-Eve satisfy the following inequality:
\begin{eqnarray}
I_{AB}^N + I_{AE}^N \leq 2 \log_2(NC).
\label{ineq1}
\end{eqnarray}

\noindent Now we are ready to prove Theorem 2.\\

{\em Proof of Theorem 2:}
Suppose Alice sent a large number of quNit symbols, and that Bob performed this measurement on $n$ quNits of them using the correct basis. The Hilbert space dimension of the total sifted symbol space is thus $N^n$. Let us now re-label the bases for each of the $n$ quNits such that, by definition, Alice used all $n$ times the $\{\psi_B\}$ basis. Hence, using this re-labeling, Bob's observable is the n-time tensor product 
$\hat{B}_1 \otimes ...\otimes \hat{B}_n$. Since Eve had no way to know the correct bases, her optimal information on the correct ones is precisely the same as her optimal information on the incorrect ones. Hence, one can bound her information assuming she measures $\hat{E}_1 \otimes ... \otimes \hat{E}_n$ where $\hat{E}_i$ is a complementary observable to $\hat{B}_i$. It follows that $C = N^{-n/2}$. By applying Theorem 3, we obtain the following inequality:
\begin{eqnarray}
I_{AB}^N + I_{AE}^N \leq n \log_2(N)
\label{ineq2}
\end{eqnarray}
By using the inequality $I_{AB}^N \geq I_{AE}^N$ of  Theorem 1 and Eq. (\ref{ineq2}) and we obtain:
\begin{eqnarray}
I_{AB}^N \leq \frac{n}{2} \log_2(N) .
\label{ineq3}
\end{eqnarray}
For  string of $n$ symbols, the mutual information between Alice and Bob becomes 
\begin{eqnarray}
I_{AB}^N(e^N_B)= n(\log(N) + (1-{e^N_{B}})\log(1-e^N_{B}) + e^N_{B}\log(\frac{e^N_{B}}{N-1})) .
\label{IABn}
\end{eqnarray}
Using the Eqs. (\ref{ineq3}) and (\ref{IABn}), we obtain Theorem 2.

In Fig. \ref{fig:EBob}, we plot the upper bound for the Bob's error rate as function of $N$ in the case of the optimal incoherent and coherent eavesdropping attacks. For $N=2$ we recover the results  for coherent attacks the result by Mayers \cite{Mayers1,Shor} $e_{B}^2 = 11\%$.

%----------------------------------------------------------------
\section{Realistic Systems}
The attacks presented in the previous sections assume perfect eavesdropping and measurement apparata, and a noise-free channel. In real secret key distribution systems there are several limitations: The sources can emit more than one photon, some photons never  
get to Bob's detector (channel loss), the detector quantum efficiency is limited, and the dark count probability (counts not produced by photons) of the detectors is not negligible. We therefore define an experimental Quantum Bit Error Rate (QBER) for a $N$-dimensional Hilbert space where we assume that the optical noise remains negligible even for large $N$ and the only source of noise is the dark count of the detectors. Under these assumptions the $QBER$ is given by:
\begin{eqnarray}
QBER = \frac{ P_{incorrect}}{P_{incorrect}+ P_{correct}} \approx  \frac{ P_{incorrect}}{P_{correct}} ,
\label{QBER}
\end{eqnarray}
where 
\begin{eqnarray}
P_{correct}= \mu \eta_D e^{-\alpha L}  \frac{1}{M}.
\label{correct}
\end{eqnarray}
In Eq. (\ref{correct}), $\mu$ is the average photon number per symbol,  $\eta$ is the detector quantum efficiency, $\alpha$ is the channel attenuation coefficient, $L$ is the transmission length and $q_{basis} = 1/M$ is a factor which depends inversely on the number of bases used in the protocol . The  probability of incorrect counts, when we assume that all incorrect counts come from the detectors and they have the same dark count probability, is given by:
\begin{eqnarray}
P_{incorrect} \approx P_{dark}(N-1)\frac{1}{M} ,
\label{incorrect}
\end{eqnarray}
where $P_{dark}$ is the probability of dark counts by detector. The 
$QBER$ becomes
\begin{eqnarray}
QBER \approx  \frac{P_{dark}(N-1)}{\mu \eta e^{-\alpha L}  } .
\label{QBERf}
\end{eqnarray}
In Fig. \ref{fig:IABEXP} we plot the information rate  $R_{AB}$  as function of the transmission distance. The intersection of the two curves gives the maximal distance allowed between Alice and Bob such they can establish a secret key with typical parameter values
$\eta_D = 20$ $\% $, $\alpha = 0.2$ $dB/km$, $\mu = 0.1$ and $ P_{dark} = 10^{-5}$. 
\par
%Error correction and privacy amplification

%==================================================================
\section{Discussion and Conclusions}

In this work we have considered an extension of Bennett's and Brassard's seminal quantum key distribution protocol into a $N$-dimensional Hilbert space. We have obtained bounds on Bob's permissible error rate in the case of individual and coherent eavesdropping attacks, and we have give the limits for the transmission distances in non-ideal systems.
Using similar arguments and methods one could also generalize Ekert's quantum cryptographic protocol \cite{Ekert}, based on quantum entanglement and the test of Bell inequality to detect the eavesdropping,  to a $N$-dimensional Hilbert space. Recently Kaszlilowski \textit{et al.} have shown \cite{zeilinger} that the violation of local realism by two entangled quNits is stronger than the violation for two entangled qubits. We conjecture that this would also imply a higher degree of security in entanglement-based multi-level quantum cryptography.

%---------------------------------------------------  
\section*{Acknowledgments}
%\begin{Acknowledgments}
We would like to thank Hugo Zbinden for useful discussions. This work was supported by the Swedish Research Council 
for Engineering Sciences (TFR), the European Commission through the IST FET QIPC QuComm and EQUIP projects. One of us, M.B., thank GAP-optique for their kind hospitality during his stay there, and the European Science Foundation for financial support.
%\end{acknowledgments}

%\vspace{-0.5cm}

\newpage

\begin{figure}[h]
\caption{The information  rate $R_{AB}$, defined by Eq. (\ref{RAB}), as a function of Bob's error rate $e_{B}^{N}$ for different  Hilbert space dimensions $N$, assuming that $M=N+1$, using the Universal Quantum Cloning Machine eavesdropping strategy.}
\label{fig:IABN}
\end{figure}

\begin{figure}[h]
\caption{Bob's error rate $e_{B}^{N}$ as a function of the dimension of Hilbert space $N$ for optimal incoherent and coherent eavesdropping strategies.}
\label{fig:EBob}
\end{figure}

\begin{figure}[h]
\caption{The information rate  $R_{AB}$, defined by Eq. (\ref{RAB}),  as a function of the transmission distance $L$ [the distance $L$ is related to the quantum bit error rate by Eq. (\ref{QBERf})]. Curves are plotted for different dimensions of the Hilbert space $N$, assuming that  $M=N+1$ and that the Universal Cloning Machine eavesdropping strategy is used.}
\label{fig:IABEXP}
\end{figure}

%\end{multicols}{2}
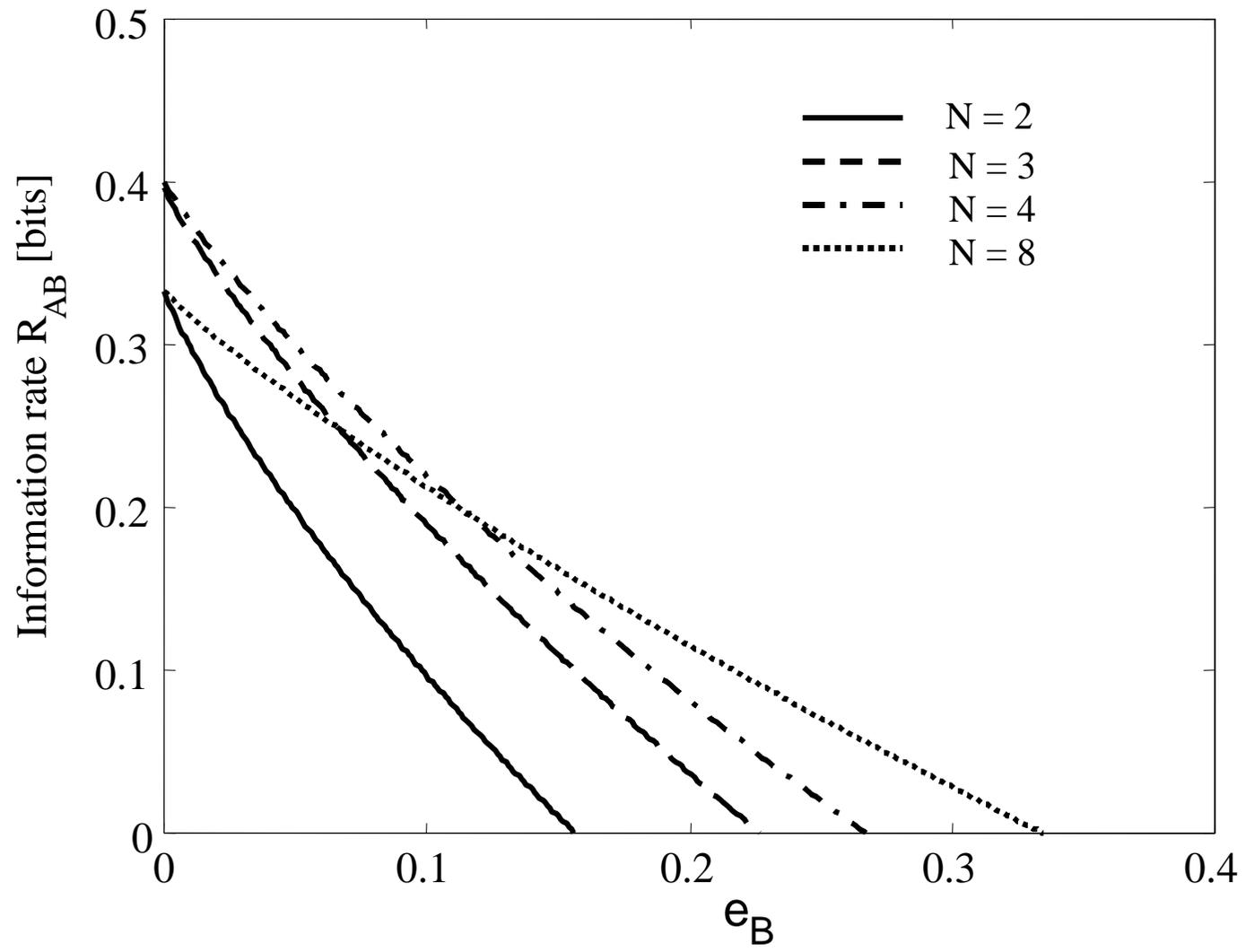
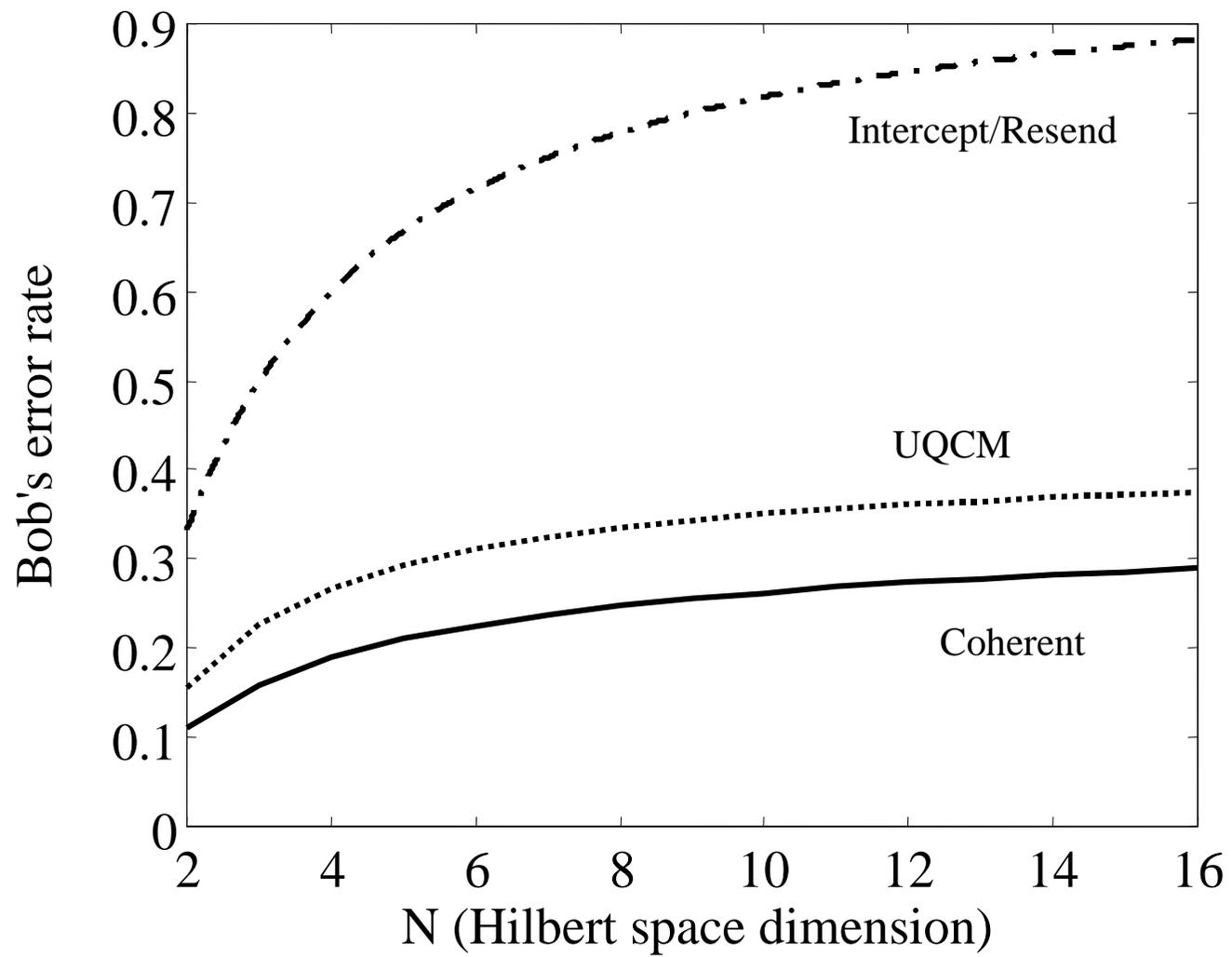
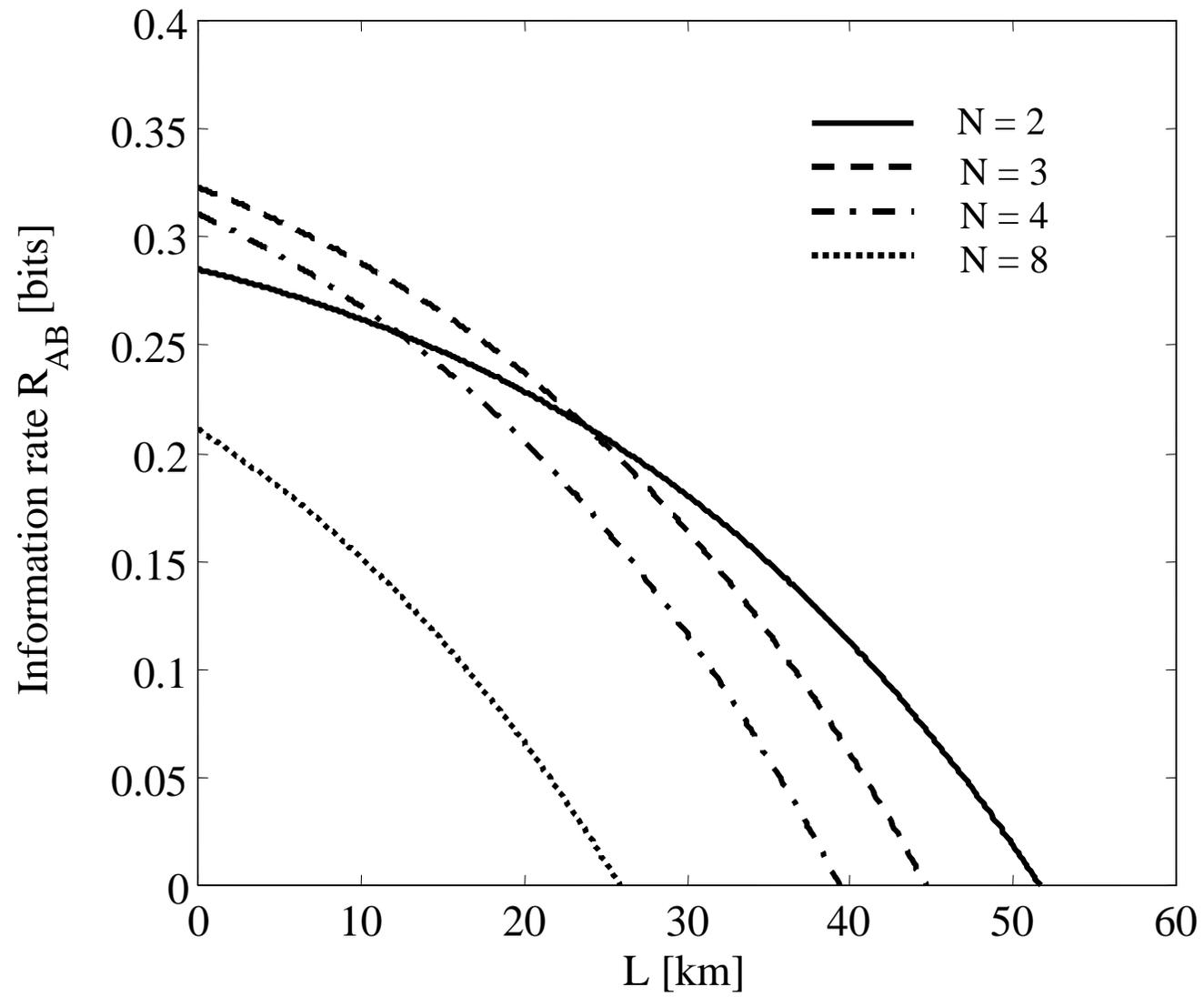
\end{document}